\documentclass[12pt,a4paper]{article}

\usepackage[british]{babel}

\usepackage[a4paper,top=2cm,bottom=2cm,left=2.5cm,right=2.5cm,marginparwidth=1.75cm]{geometry}

\usepackage[style=ieee, backend=biber]{biblatex} % APA 7th edition style citations using biblatex
\DeclareLanguageMapping{british}{british-apa} % Set language mapping
\usepackage{amsmath,amssymb,amsfonts}
\usepackage{graphicx}
\usepackage[colorlinks=true, allcolors=blue]{hyperref}
\usepackage{hyperref}
\usepackage{orcidlink}
\usepackage[title]{appendix}
\usepackage{mathrsfs}
\usepackage{amsfonts}
\usepackage{multirow, tabularx, booktabs}
\usepackage{caption}  % For \caption
\usepackage{threeparttable} % For table footnotes
\usepackage{algorithm}
\usepackage{algorithmicx}
\usepackage{algpseudocode}
\usepackage{listings}
\usepackage{enumitem}
\usepackage{chngcntr}
\usepackage{booktabs}
\usepackage{lipsum}
\usepackage{subcaption}
\usepackage{authblk}
\usepackage[T1]{fontenc}    % Font encoding
\usepackage{csquotes}       % Include csquotes
\usepackage{diagbox}
\usepackage{adjustbox}

\AtEveryBibitem{%
  \clearfield{url}
  \clearfield{issn}
  \clearfield{note}
  \clearfield{eprint}
  \clearfield{month}
  \clearfield{pages}
  \clearfield{day}
  \clearfield{issue}
  \clearfield{volume}
   \clearfield{number}
  
  \ifentrytype{article}{%
    \clearlist{publisher}  % Remove Publisher only for journal articles
  }{%
  }
}

\usepackage{setspace}
\usepackage{titlesec}
\titleformat{\section} % Redefine section numbering format
  {\normalfont\Large\bfseries}{\thesection.}{1em}{}
  
\usepackage{lineno} % Add the lineno package
\usepackage{float}   % for customizing caption position
\usepackage{caption} % for customizing caption format


\title{The Costs of Competition in Distributing Scarce Research Funds}

\author[1,*]{Gerald Schweiger}
\author[2]{Adrian Barnett}
\author[3]{Peter van den Besselaar}
\author[4]{Lutz Bornmann}
\author[5]{Andreas De Block}
\author[6]{John P.A. Ioannidis}
\author[7]{Ulf Sandström}
\author[8]{Stijn Conix}

\affil[1]{\small Institute of Software Technology, Graz University of Technology, Graz, Austria}

\affil[2]{Australian Centre for Health Services Innovation and Centre for Healthcare Transformation, School of Public Health and Social Work, Faculty of Health, Queensland University of Technology, Kelvin Grove, Australia}
\affil[3]{Department of Organization Sciences and Network Institute, VU University Amsterdam, Netherlands, and DZHW - German Centre for Higher Education Research and Science Studies GmbH, Berlin, Germany}
\affil[4]{Science Policy and Strategy Department, Administrative Headquarters of the Max Planck Society, Munich, Germany}
\affil[5]{ Center for Logic and Philosophy of Science (CLPS), Institute of Philosophy, KU Leuven, Leuven, Belgium}
\affil[6]{Departments of Medicine and of Epidemiology and Population Health and Meta-Research Innovation Center at Stanford (METRICS), Stanford University, Stanford, USA}
\affil[7]{Department of Industrial Economics and Management, KTH Royal Institute of Technology, Stockholm, Sweden}
\affil[8]{Institut supérieur de philosophie, Université catholique de Louvain, Louvain-La-Neuve, Belgium}

\affil[*]{Corresponding author: \texttt{schweiger.gerald@gmail.com}}

\date{}  % Remove date

\begin{document}
\maketitle

\begin{abstract}
Research funding systems are not isolated systems – they are embedded in a larger scientific system with an enormous influence on the system. This paper aims to analyze the allocation of competitive research funding from different perspectives: How reliable are decision processes for funding? What are the economic costs of competitive funding? How does competition for funds affect doing risky research? How do competitive funding environments affect scientists themselves, and which ethical issues must be considered? We attempt to identify gaps in our knowledge of research funding systems; we propose recommendations for policymakers and funding agencies, including empirical experiments of decision processes and the collection of data on these processes. With our recommendations we hope to contribute to developing improved ways of organizing research funding. %\lipsum[1]
\end{abstract}

%\textbf{Keywords}: keyword1, keyword2, keyword3, keyword4, keyword5, keyword6.  

%-------------------------------------------
% Paper Body
%-------------------------------------------
%--- Section ---%
\section{Introduction}\label{sec1}
Scientific research is expensive.
On average, OECD countries allocate 2.7\% of their GDP to scientific research, with some countries spending up to 5\% \cite{OECD23}.
Given this substantial social commitment to the funding of science and the profound impact of science on society, it is imperative that funding is allocated efficiently \cite{ioannidis2011}. Society expects only research that offers the greatest social benefits in terms of economy, health, culture, etc., to receive funding, although there is considerable disagreement with regard to defining these “greatest social benefits” and determining the criteria for allocation in a democratic society \cite{kitcher2001science}.
Despite massive investments in science worldwide, rigorous scholarly investigations into science funding had long been very limited, and studies that evaluate and compare alternative funding systems (besides mere descriptive approaches) have been scarce.
Over the past decade, however, a growing field has emerged that is starting to address this gap \cite{shaw2023a, fang2016a, shaw2024bias, bollen2017}.
Determining the best ways to distribute research funding remains challenging, arising primarily from the inherent complexity of the scientific system (and the funding system), that is characterized by nonlinearity and multidimensionality \cite{Latour1987,shi2015}. Both characteristics make it challenging to identify and test causal connections. Performing experiments related to science funding is likely to be resource-intensive and require extended time frames, making it challenging to obtain experimental data.
%Consequently, researchers often resort to drawing causal inferences from non-experimental data, demanding an \textit{"enormous investment of skill, intelligence, and hard work"} \cite{Freedman2005}.

The lack of consensus on the precise aims of research funding compounds this complexity. 
While many policymakers and scientists may argue that scientific ``excellence'' should be funded, ``excellence'' is an ambiguous term, and its use has been viewed increasingly with suspicion \cite{merton1973sociology, moore2017excellence}.
Some view its vagueness as beneficial, fostering collaboration among diverse actors with distinct perceptions of ``excellence'' \cite{hellstrom2011homing}. Others interpret ``excellence'' as counter-productive by potentially playing a purely rhetorical role without any valid and operational meaning necessary to shape and guide research funding \cite{o2020micro}. 
Even if funding systems use ``excellence'' as a general criterion, its precise meaning must be defined in relation to the many different aims funding instruments may have: contributing to a specific focus area; societal and economic impact; career support to particular categories of researchers to become independent (e.g., early-career researchers, women, underrepresented minorities); fostering internationalization; supporting interdisciplinarity, etc. It is crucial to be specific about the various aims of science funding and the criteria that follow, and it is as crucial to be specific on how to measure these criteria so as to avoid noise and bias \cite{kahneman2021noise}.
In the absence of a clear definition, ``excellence'' is often operationalized using bibliometrics such as citation counts, without taking into account that these cannot cover all aspects of research quality and impact \cite{ioannidis2023defense,van2019measuring}.
While experts in the field of bibliometrics emphasize that citation data can capture only one aspect of quality - research impact (and even this may only be captured imperfectly) \cite{hicks2015bibliometrics} - citizen (or lay) bibliometricians frequently use bibliometrics as a surrogate for all dimensions of research quality \cite{leydesdorff2016professional}. The issue of ``excellence'' indicates that not only an empirical but also a conceptual analysis of the allocation of research funds is crucial to optimize the system.

\bigbreak
 This paper aims to contribute to this goal by analyzing research funding from economic, epistemic, social, and ethical perspectives. We also discuss open research questions that should be empirically answered. This paper refrains from delving into high-level decision-making processes, including fundamental decisions such as the shares of basic and applied research or the distribution of funds over the various disciplines. Although these high-level science policy decisions are important, they require different considerations.

\section{Competitive funding }\label{sec2}
At a high level, one can distinguish between two types of distribution systems: Either scientists actively compete for research funds (e.g., by writing research proposals) or they receive funds without actively doing so (e.g., through direct block funding of universities) \cite{sandstrom2018funding}. 
%Competitiveness of a national distribution system in the context of academic funding is often defined as the share of project funding (as opposed to basic university funds) in total research funding, \cite{abramo2012dispersion}.
%There is no consensus in the literature on the definition of the term competition. However, we can establish some minimum conditions that must be met if something is to be classified as competitive \cite{Altreiter2023three}.
%According to Arora-Jonsson et~al. \cite{arora2021competition}, one of these minimum requirements is that at least two actors (who may be individuals, groups or other entities) compete for the same scarce resources. These do not necessarily have to be monetary resources; for example, reputation can be seen as a resource that is competed for. 
%Furthermore, there needs to be an allocation mechanism that distributes these scarce resources based on certain criteria, e.g., scoring applications using peer review.
Central to the development of competition in the academia was the move towards New Public Management, which refers to various reforms aimed at increasing the efficiency and performance of public organizations \cite{broucker2015new,krucken2021multiple}.
Academic competition occurs at different levels (individual, institutional, national) and among various actors \cite{krucken2021multiple,musselin2018new}.
%This paper discusses competition in the allocation of research funds: individual researchers or collaborative research efforts compete with others for scarce research funds. 
Comparing countries reveals a significant variation in the balance between institutional block funding and competitive project funding, influencing the level and nature of competition for research funding \cite{sandstrom2018funding}. Block funding also flows from national research evaluation systems, further fueling competition. Also, block funding can be distributed competitively within the universities \cite{auranen2010university}. 
% I don't understand the following sentence. What are the 'instruments'? Is that peer review systems? If so ``may also be different'' should probably be ``unknown'' or ``untested''.
Even within competitive grant allocation, many different instruments exist in which the costs and benefits balance may also be different \cite{Thelwall2023}.     
The balance between these various forms of funding may be decisive for the relation between the costs and benefits of competitive research funding.

\subsection{Decision-making processes 
}\label{sub_dec}

Grant peer review operates through two primary models: independent reviews followed by panel discussions and standing panels with dedicated members \cite{mitroff1979peer, mccullough1989first}. Reviewers are asked to follow predefined criteria set by funding agencies, adapted to submission types \cite{bornmann2011scientific}. Panel peer reviews may include meetings for discussion and consensus-building on borderline cases. After completing the review process, the funding agency processes evaluations and recommendations for the final decision on acceptance, revision, or rejection.
In the ex-ante evaluation of project ideas, a significant irrational element exists inherent in the heavy reliance on a small group of experts to screen, rank, and select among potentially promising research projects.
This is because peers, in principle, cannot fully discern the true potential behind written proposals. The limitation lies in the subjective nature of evaluation, where experts may not always grasp a project's full scope or potential solely from its written description \cite{roumbanis2021oracles}.
Previous studies have indicated that the grant review process has several weaknesses related to issues such as conflicting interests \cite{sandstrom2008persistent, breen1997nepotism}, promoting conformity \cite{bakanic1987manuscript}, promoting conservatism \cite{luukkonen2012conservatism, alberts2014rescuing}, early-career blindness \cite{marsh2009gender}, and reviewers favoring applicants from their field \cite{travis1991new, wang2015defining}. Furthermore, studies show that reviewers who are applicants themselves are more successful than applicants with better bibliometric performance \cite{van2012selection, abrams1991predictive}.

Most of the identified weaknesses address three fundamental aspects: the reliability, fairness, and predictive validity of peer review. 
Reliability is usually examined on the basis of agreement between reviewers. 
Some studies show no agreement when comparing scores between reviewers of the same grant proposal, and authors highlight the subjectivity of reviewers’ assessments, concluding that the peer review process is arbitrary \cite{Pier2018a}. Other studies show very low \cite{mayo2006,mutz2012} to low agreement \cite{erosheva2021}. Since access to rejected proposals is frequently restricted, some of these studies are statistically flawed as they only analyze accepted proposals \cite{erosheva2021}. 
Findings on the reliability of peer review in paper evaluation for journals confirm the results for grant peer review \cite{bornmann2010reliability}.  
Training peer reviewers for journals and grant reviews leads to little or no improvement in the quality of peer review \cite{hesselberg2023reviewer}.
The reliability of panel decisions is usually examined based on agreement between independent panels.
The presence of chance in panel decisions was confirmed in quantitative \cite{cole1981, graves2011} and qualitative studies \cite{lamont2009, roumbanis2017, Coveney2017}.

\bigbreak

Results concerning the predictive validity of grant peer review show no \cite{fang2016a,danthi2014, danthi2015, van2015, doyle2015, kaltman2014} or a weak relationship \cite{gallo2014,lauer2015, reinhart2009} between ranking percentile after peer review and scholarly success measured by bibliometric indicators. In one study, researchers with awarded projects and researchers with rejected projects in the biological and social sciences were compared based on bibliometric indicators \cite{Bornmann2010}. If only the best rejected applicants (same number as awarded) are considered in both fields, they score better than the awardees on citation impact. A review suggests that peer review-based decision processes may be able to separate good and flawed proposals, but discrimination amongst the top tier proposals or applicants may be more difficult \cite{gallo2018external}. The best rejected proposals score on average as high on bibliometric indicators as the accepted proposals \cite{Thorngate2002}. 
%Past performance shows slightly better results than peer review of research projects \cite{kaltman2014}. 

\bigbreak

If applications are not only assessed based on universal criteria (the quality of research) following the ethos of science \cite{merton1973sociology}, but on personal criteria (such as gender or nationality), the fairness (and validity) of the peer review process is compromised.
Many personal criteria have been investigated in the context of peer review processes, with most studies examining applicants' gender. The historical perspective suggests that gender bias in review scores and grant decisions has declined considerably over the last decades. 
Recent reviews found no significant differences in grant success between women and men after controlling for performance and other relevant covariates, but women received smaller awards and fewer awards after re-applying \cite{schmaling2023gender, kahn2023there}. 
It should be noted here that non-merit criteria can be included in application evaluation criteria, as grants may, for example, aim to support the academic careers of key groups, such as grants for early-career researchers or for women in fields where women are underrepresented.

\bigbreak

\section{Costs of competition }\label{sec3}

\subsection{Economic costs of competition 
}\label{subsec31}

The economic costs of competitive funding systems include the costs of applicants needing time to write proposals, decision-making processes (reviews and panels), and administrative costs. A study in Australia estimated that 85\% of the costs are incurred by the applicants, 10\%  by the decision-making processes, and the remaining 5\% by the administration \cite{graves2011}. Funding schemes will have a net financial gain of zero if the costs for these three categories are equal to the amount of funds awarded \cite{Dresler2023}; this is most likely for schemes with many applicants and low success rates.
Depending on the total available funding, the funding rate, and the time that the applicants invest in the proposals, the point of zero net gain is crossed at funding rates that are not uncommon in current funding systems \cite{Dresler2023,barnett2021a}.

Studies in various disciplines show that writing a single proposal takes about 25 to 50~days \cite{schweiger2023, von2015, herbert2013}. With average acceptance rates between 10 to 25\%, that is 100 to 500~person-days of effort per funded project. In interpreting reported studies on time investment, it should be considered that they are primarily based on self-reporting, and self-reporting of time use may be inaccurate in many situations \cite{stone1999science}.
%A study analyzing funding for medical research in Australia shows that in 2013, an estimated 550~working years of researchers' time were spent preparing proposals; this represents €41~million in salaries and 14\% of the total medical research budget \cite{herbert2013}. 
An evaluation of Europe's H2020 programs found that between 30 and 50\% of the funding from Horizon~2020 is spent on grant writing \cite{European-University-Association2017}. 
A study of federally funded research projects in the U.S.\ estimated that principal investigators spend on average about 45\% of their time on administrative activities related to applying for and managing projects rather than conducting active research \cite{Schneider2020}. 
A time allocation study of academics in the US estimated that they dedicate an average of 4.6~hours per week to grant writing, with already tenured academics spending less time \cite{Link2008}.

There are limited empirical results on the time reviewers spend reviewing applications.
An evaluation of the UK Research Council estimated that reviewers spent about 192~person-years reviewing applications \cite{Vaesen2017}. 
Feedback from reviewers does not seem to benefit the quality of the proposed research \cite{herbert2013}: A study found that resubmitted proposals had a lower probability of winning funding than proposals submitted for the first time \cite{herbert2013}.

Success rates in private funding schemes are often even lower than for academic grants.
A funding program for women in STEM has a success rate of less than 1\%; only 6~out of 650 applicants were awarded funding \cite{barnett2021a}. If applicants spend more than 7~days to complete their application, the point of net zero gain is crossed.
In a private research call in computer science, 2~out of 1090 applications were funded \cite{mozilla}. With a total prize of \$120,000, net zero gain is crossed if each application costs just \$110.

\bigbreak
Writing funding applications may also have positive indirect effects. Scientists can generate, refine, and share their ideas, regardless of whether or not they ultimately receive the funding \cite{Myers2022}. 
Furthermore, even if the overall costs of the competitive funding are equal to the amount funded, at the individual level, there is a financial benefit for those who win the grants at the expense of those who do not. And the research done could theoretically be better than without the competition.
Since sharing and refining research ideas is not only restricted to environments with competitive funding systems, the size of the indirect effect is not clear. One could imagine that scientists are more willing to share ideas, data, and results in an environment with less competition.

\subsection{Epistemic costs of competition 
}\label{subsec4}

Research funding has multiple epistemic aims, including promoting groundbreaking and high-risk research \cite{wang2017bias,wu2019large}, supporting incremental progress within established paradigms \cite{thomas1962structure}, and translating theoretical insights into practical applications. The impact of peer review and competitive funding on achieving these different objectives will vary, and an ideal funding strategy should tailor its distribution methods to the specific epistemic goals of choice. The plurality of epistemic choices makes it difficult to assess the epistemic costs of competitive funding. In this section, we discuss those epistemic costs that are most obvious and have been discussed in the literature.

Most prominent in the literature is the supposed relationship between competitive funding and the lack of high-risk, high-impact research. Empirical studies indicate that major scientific breakthroughs are infrequent \cite{wang2021science} and the distribution of research outcomes is positively skewed, with highly productive outcomes being rarity \cite{radicchi2008universality}. While such outcomes would likely be rare under any funding method, it has been argued that peer review of grant applications deters researchers from proposing high-risk research. On the one hand, writing and reviewing applications reduce the time for such research, which is probably more time-intensive than conventional research. On the other hand, a study of the European Research Council's funding decisions showed a bias against applicants with histories of high-risk research, favoring those with more conventional profiles \cite{veugelers2022funding}. The main bias against applicants with very high-risk and, thus, controversial profiles occurs in the first stage of selection when panel members review applications based on a brief description of the proposed research and the applicant's CV. Other studies draw similar conclusions that high-risk research is disadvantaged in competitive funding systems \cite{lanoe2018evaluation,boudreau2016looking,wagner2013evaluating}. However, more empirical studies are required to confirm and explain this finding.

Researchers might adjust their behavior in response to the perceived bias against high-risk proposals. A theoretical study suggested that the predominance of conservative submissions is larger in grant proposals than in paper submissions. This is because grant proposals need some preliminary promising data to support the hypotheses, which might lead peer reviewers to favor safer ideas \cite{gross2021ex}. A study in life sciences found that programs that reward long-term success lead to higher levels of breakthrough innovation arising from risky research, compared with short grant cycles and predefined deliverables \cite{azoulay2011incentives}. Ideas for high-risk research need time, and their implementation requires time and a dependable environment. 
%Another study suggested that in ERC schemes that claim to support high-risk/high-gain proposals, such proposals have, in fact, a lower probability of getting funded \cite{veugelers2022funding}. 

The competitive nature of a system might affect not just the proportion of risky research but also researchers' performance, as indicated by bibliometric indices. A study analyzed the influence of competitive research funding on efficiency in 17~countries, defined as the change in funding compared with the change in highly cited publications \cite{sandstrom2018funding}. The data show a negative correlation of 0.3 between efficiency and the degree of competitive funding. In another study, data from eight countries were analyzed to assess the impact of competitive research funding on productivity as measured by the number of publications  \cite{auranen2010university}. Countries with high competition (e.g., the UK) are efficient but have not been able to improve efficiency; other countries with less competition (e.g., Denmark) are either almost as efficient or (e.g., Sweden) have been able to improve their efficiency despite relatively low competition.
A study comparing competitive and institutional block funding within Japan reveals mixed results concerning the novelty of outputs measured using bibliometric methods. Competitive funding appears to be associated with higher novelty compared with papers funded by block funding. However, these findings are contingent upon the status of the researchers. Specifically, high-status, senior, and male researchers tend to produce more novel outputs under competitive funding schemes. Conversely, lower-status, early-career, and female researchers had a negative relationship with novelty. These nuanced results underscore the complex interplay between funding mechanisms and researcher characteristics in shaping research outputs \cite{wang2018funding}.
Note, however, that the latter three studies are not designed to enable the discovery of causal relationships. Further research is needed to uncover the underlying mechanisms.

If the conservatism of peer reviewers contributes to competitive funding’s bias against highly novel, risky research, then such funding might be more appropriate for funding incremental scientific progress. Yet, even here, excessive competition may incur epistemic costs, particularly in research communities tackling complex topics. A simulation study coupled with historical data shows that such research communities tackling complex topics (e.g., fundamental problems in physics) are at risk of holding on to popular, but incorrect paradigms due to low self-correction capabilities \cite{fang2011peer, liu2012peer}. Science funding systems should facilitate the quick abandonment of fruitless ideas and minimize investment in scientific dead ends. However, this may not happen since low-yield and fruitless ideas are still defended by communities of scientists and organizations who have made careers based on them \cite{joyner2016happens}. Conversely, successful groups should be protected from collapsing due to funding gaps. 

Do the top-cited scientists receive a steady stream of funding? A study analyzing the association between federal biomedical funding and high citation impact among scientists found that while funded top-cited scientists attracted more citations than non-funded, only a small minority had funding at the time of the survey \cite{ioannidis2022federal}.
A study in the UK in the field of health research came to similar conclusions. The majority of the UK's most influential health scientists do not receive funding from the country's top three public and charity funders \cite{stavropoulou2019most}. 
This research must also be conducted for other disciplines and countries.
If only a limited proportion of top-cited researchers currently receive funding from public funders, this must be considered in thoroughly evaluating existing funding systems.

Competitive funding also incurs epistemic costs that probably affect all epistemic aims of science. For instance, the economic costs discussed above have a direct epistemic counterpart: the time spent by researchers on writing and reviewing proposals consumes a portion of the total research budget. Economically more efficient ways of distributing the funding would translate to additional epistemic payoff. Even if funded projects in a competitive system perform better than rejected ones, it must be demonstrated that this outweighs the resources that could be allocated to additional research in alternative systems. 
%However, reviewing, just like proposal writing may have learning effects for those involved, and studies suggest that this is indeed recognized by reviewers \cite{Van_den_besselaar_AvH}.

Another epistemic cost of competitive funding relates to biases, particularly cronyism, in competitive funding. Diversity in background assumptions, perspectives, and research interests is vital for research communities to remain critical and self-correcting \cite{longino1990science}. If competitive funding leads to a homogenization of viewpoints, whether through biases or conservatism, research communities dependent on such funding are more susceptible to confirmation bias and blind spots that would more likely be corrected in more diverse communities.

\subsection{Social and ethical costs of competition 
}\label{subsec5}

Some of the epistemic and economic costs discussed above also come with distinct social and ethical costs for the society as a whole or certain groups within the society. 
%Social costs should be avoided, especially in science funding, since a large proportion of research funding is drawn from societal funds. 
Inefficient allocation of societal resources comes at the expense of opportunities, as the probably wasted resources could be spent on direct improvements of well-being such as medical care, poverty relief, and other pressing societal problems. 
%Potential cronyism, gender bias, and other biases in science funding processes may create a substantial social cost as they discriminate against groups that already face discrimination in many other domains. But this may also play a role in the non-competitive allocation of research money.
Competitive funding processes create distinct groups of winners and losers. The Matthew effect leads to the manifestation and reinforcement of winning and losing in the processes. Most funding schemes have low success rates under 25\%, and in many cases even under 15\% \cite{Crew2019}, so failure is the norm. While the fortunate winners can continue their scientific careers, the losers can face big decisions, including changing their research plans, relocating to another city or country, or even deciding to quit research \cite{conroy2020}. The social costs can extend beyond the scientist to their family \cite{herbert2013}, as bluntly described by one researcher: ``My family hates my profession. Not just my partner and children but my parents and siblings. The insecurity despite the crushing hours is a soul destroying combination that is not sustainable.'' \cite{herbert2014}. Failing to win funding has been compared with the grief of bereavement \cite{borgstrom2023grieving}.

The pain of failure is often compounded by regret as researchers reflect on the opportunity costs of the time spent writing failed applications that could have been spent on actual research. 
Whereas we know from journal peer review that many rejected manuscripts are submitted elsewhere (the time spent on writing the manuscript was not wasted), this does not appear to be the case with rejected proposals: The results of a previous study in medical research show that resubmissions take only 25\% less time than submitting a new application \cite{herbert2013}; the result of a study in engineering showed no differences between the time needed to write new applications compared with resubmissions \cite{schweiger2023}.

An estimated 90\% of researchers perceive that they spend too much time preparing competitive research proposals, and only 10\% of researchers believe that the current competitive funding system positively affects the quality of research \cite{schweiger2023}. 

The immense pressure to win funding can negatively impact researchers' mental health and work-life balance \cite{Hall2023,woolston2020postdoc}. Some researchers even choose to have fewer children to remain competitive \cite{Ecklund2011}. For those with children, there is an inevitable disruption to their career, particularly for female researchers who are often the primary caregivers for children and relatives \cite{Cech2019}. This means women often have less time to apply for funding, which may partly explain the common male-female funding gap \cite{Kingsley2023}. 
Many funders have recognized the impact of caring for children on researchers' careers and allow applicants to explain how any disruption has impacted their productivity, aiming to level the playing field. However, this may not always work as women have reported not wanting to document their disruption for fear of appearing ``weak'' \cite{Barnett2022,Pribbenow2010}. It is unclear how reviewers actually respond to such disclosures and whether disclosures level the playing field or make inequities more prominent.
Furthermore, hyper-competitive and pressurized funding systems can reduce collegiality, sow distrust in the community, and create feelings of alienation \cite{Anderson2007,Osmond1983}. 

Although these criticisms of competition have credibility, they do not mean that less competitive funding and more block funding would solve these problems. Competition is inherent in science (besides cooperation as the other side of the coin). It is not restricted to grants but is also true of academic positions, promotion to higher positions, etc. Even if all funding were awarded as block funding, there would still be not enough research funding to create a sufficient research environment for every staff member. Decisions about scarce resources would become bureaucratic (and potentially more biased), and the competition would shift to colleagues from the same organization vying for a larger slice of the block funding. It is doubtful that this would improve the quality of research and reduce social costs.      

There is also a moral dimension to many of the problems listed above: discriminating based on gender, age, and race is morally wrong, as is wasting societal funds through their inefficient allocation. In addition to these moral problems, competitive grant funding may also incentivize direct violations of core values of research integrity such as honesty, accountability, fairness, impartiality, and responsibility \cite{Conix2021}. This seems plausible because success in competitive funding carries substantial weight in tenure, hiring, and promotion decisions \cite{schimanski2018evaluation}. In combination with low success rates and researchers’ reliance on grant money for research, applicants face strong incentives to cut corners in the application process that might improve their chances \cite{Conix2021}. For example, competitive grant funding might incentivize practices such as salami-slicing and other publication and authorship-related questionable research practices (QRPs), as a researcher’s track record is often an important factor in evaluating grant proposals. These incentives are further exacerbated by funders prioritizing ex-ante grant peer review over post hoc project evaluation, meaning that many violations will go unnoticed \cite{gopalakrishna2022prevalence}. Indeed, one study found evidence for double dipping \cite{garner2013same}, and a recent survey found that a significant portion of respondents admitted to frequently engaging in various QRPs, including improper use of funds, overstating confidence in the research proposal, double dipping, and selective citing \cite{garner2013same,conix2023questionable}. In the same survey, reviewers and panelists also indicated that they have frequently engaged in funding-related QRPs, including inadequate preparation for panel meetings or reviewing proposals of close colleagues and friends.

There are other potential moral problems with competitive funding, but research is mostly lacking. For example, various authors hypothesize that extreme competition encourages gaming, commonly known as ``grantsmanship'', such as writing the application you think will get funded rather than the work you think is most important \cite{Anderson2007,Osmond1983}. Other potential harms include actively avoiding collaborations with peers so as to have an appropriate expert to review your applications \cite{barnett2023}. 
Some researchers do not trust funding peer review systems and are concerned that unscrupulous reviewers will steal their ideas \cite{Conix2021}. Occasionally, researchers move beyond gaming and into outright dishonesty \cite{Anderson2007,Conix2021}, and the desire to win funding has motivated some researchers to commit fraud by faking data \cite{CCC2017,ORI2022}. 
The perceived excessive costs of writing lengthy applications could drive researchers to use large language models to write parts of their applications, with ethical consequences \cite{parrilla2023chatgpt}. 

More research is needed on the moral problems related to competitive research funding, as almost all evidence is drawn from surveys using convenience samples. In addition, while some researchers may engage in QRPs intentionally to bolster their funding prospects, these QRPs are probably also ingrained in a research culture that junior researchers observe and imitate. In that sense, a targeted attempt to change the culture of research integrity within the funding context could address these moral problems \cite{gopalakrishna2022prevalence}. However, institutions may be hesitant to promote such changes, as they tend to benefit from success in funding competitions. In addition, many funding-related QRPs are intricately linked to peer review and competitive grant distribution, making it hard to eliminate them without changing the distribution model \cite{Conix2021,conix2023questionable}.

\section{Road to progress}\label{sec13}

In the previous section, we have outlined several prevalent costs entailed by competitive funding allocations. This section introduces a series of open research inquiries crucial for optimizing the allocation processes. Some issues can be assessed through measurable means, such as controlled experiments (e.g., assessing the reliability of peer review and panel decisions) or derived from data (e.g., evaluating application writing costs). However, some challenges can only be minimally measured, and controlled experiments are impractical or impossible due to system complexity or long-term effects. 
In such situations, researchers have frequently conducted studies examining the funding ecology instead of experimentally investigating individual funding schemes or organizations \cite{sandstrom2018funding, auranen2010university}. 
To move forward, a combination of studies using different designs is needed: ecological studies, possibly combined with simulation models with hypothetical causal relations, causal analysis using experiments including natural experiments, and causal analysis on cross-sectional data \cite{huntington2021effect}. Together, these can be the basis for policy decisions. The open research topics that we identified are as follows.

\bigbreak

\textbf{\emph{The foundation: Research on research}}. The need to apply the scientific process to the processes of science has been acknowledged for over a century, with a 1910 article stating, \textit{``we are at present almost wantonly ignorant and careless regarding the conditions which favor or hinder scientific work"} \cite{cattell1921american}. A paper published in 1976 concluded that \textit{``because the very nature of research on research [...] requires long periods, we recommend that independent, highly competent groups be established with ample, long term support to conduct and support retrospective and prospective research on the nature of scientific discovery"} \cite{comroe1976scientific}.  
But since that 1976 statement, there has been substantial growth in research groups on the science of science, specialized conferences, and journals, learned societies, and specific funding schemes, e.g., at the National Science Foundation and as part of the European framework programs. In addition, there are also many meta-researchers who do this in addition to their jobs in other (often medical) fields. 
Some of these researchers are organized in the Research on Research Institute (RoRI) – a nonprofit community interest company founded in 2019. The company accelerates “transformative research on research systems, cultures and decision-making” 

We can think of no other industry that spends so little on evidence-based quality control and process improvement. This lack of re-investment means that funding systems have rarely been tested and have survived for decades without change. Their longevity has put them on a pedestal, and challenging them can be treated like heresy \cite{Barnett2016}.

\bigbreak

\textbf{\emph{First: Data}}. A systematic review of funding systems concluded that research funders should \textit{„build in before and after comparisons; strive to make data available for analysis; openly publish studies of their processes and work together on comparative analysis“} \cite{Guthrie2018}. 
Non-sensitive data that does not raise privacy or data protection concerns should be accessible to the public without restrictions. This includes the number of submitted grants, funding rates, and data on the difference between the requested and allotted budget.
Data on the applicant pool is also crucial. Examining potential gender bias in funding decisions necessitates knowledge of the expected number of male and female applicants. If the number of female researchers eligible to apply is limited, a correspondingly low number of positive funding decisions can be expected.
However, for most research questions, it is crucial that these data can be linked at the researcher level so that one can control for the relevant covariates when analyzing the reliability and possible bias of grant evaluation.

Some research questions require more detailed data on the applications and, in some cases, the application text itself.
Receiving the data for accepted and rejected projects is crucial for assessing the evaluation process. This information can be used to analyze whether the rejected applications were accepted elsewhere.

We would need the application data when analyzing whether specific technical terms or terms, such as ``innovative'', ``ground-breaking'', or ``novel'', occur more frequently in successful applications.
Funders could consent to applicants' data being used for research purposes when they submit their applications. It is also conceivable that some data must be made available to submit to certain calls. Researchers who provide a protocol and get ethical clearance from their institute could—with a non-disclosure agreement and an adequate data management plan—be given access to these microdata.
Another option is to use the \textit{OpenSafely} approach for research data access, where researchers submit their analytical code and, without accessing the original data, receive summary results \cite{opensafely}. 
%This process is facilitated by the data owner (e.g., a statistical office or a research funder), who conducts the analysis based on the researcher's protocols.

A general problem with data from organizations such as funding agencies is that the data can lead to criticism that strikes at the core of their practice. This can lead to defensive behavior; therefore, clear policy guidelines are needed.

\bigbreak

\textbf{\emph{Second: Reliability and predictive validity of funding decisions.}} Whereas many studies have been conducted that investigate the reliability of journal peer review processes, we need more results on the reliability of funding decisions. The few studies on the reliability of peer reviews and panel decisions have examined research tenders in basic research. Future research should try to broaden this perspective: It should be investigated whether the reliability of decision-making processes differs across disciplines and how reliable the processes are in applied and especially interdisciplinary research. Since it is easier to define in basic science which people are suitable for reviews and panels, the reliability of the process may be given to a certain extent. In applied and interdisciplinary research, the selection of reviewers is often not so easy due to the broad and unspecific disciplinary nature of the projects.
The peer review panel could be further extended in applied research by including experts from outside the science sector. 
Beyond scientific excellence, evaluation criteria frequently encompass additional factors such as market potential and a nation's contribution to technological leadership. Although these considerations are justified from social and political standpoints, they may lead to additional uncertainty in the decision-making process, especially if one does not have a clear operationalization and measurement strategy for such criteria. Without these, more criteria may only lead to more noise and bias in grant selection.

Regarding the methods used to investigate judgments and decisions, we recommend that the reliability of processes should be evaluated in controlled experiments \cite{cruz2023gender} and with methods allowing for causal analysis with cross-sectional data such as matched pairs, regression discontinuity analysis, and differences in differences. Additional studies on funding processes are also necessary for predictive validity.

\bigbreak

Most past studies used bibliometric indicators to investigate the judgments of peers and funders. The strong focus on bibliometrics is particularly inadequate for analyzing processes for funding applied research. In addition, today, funders of research (the government) expect (measurable) impact of (basic and applied) research on sectors of society beyond the science sector. This impact cannot be measured with bibliometric indicators. Societal impact may be measured by alternative indicators. The citation of scientific papers in patents, e.g., is an established indicator for measuring the economic impact of research. Another recent example of data measuring research impact beyond science is data from the Overton database \cite{szomszor2022overton}. The data can be used to measure the impact of (funded) research on the policy sector. The \textit{Overton} database includes documents from the policy sector and the cited scientific literature in these documents. 

Suppose one is interested in the contribution of research to solving societal challenges. In this case, it is possible to measure the contribution of (funded) papers to the United Nations Sustainable Development Goals (SDGs) \cite{ciarli2022changing}. The SDGs address the biggest problems facing the world today (e.g., climate change or poverty). Various approaches have been developed to match published research to at least one SDG. One should nevertheless beware that we know far more about the limitations, caveats, and gaming potential of traditional bibliometrics than those of alternative indicators. Alternative indicators may be as gameable or even more gameable than traditional bibliometrics. Before we use alternatives to bibliometrics, it is necessary to know the strengths and weaknesses of the metrics.

\bigbreak

\textbf{\emph{Third: Alternative evaluation systems.}} 
 If peer review were a drug, it wouldn't be allowed on the market because it has not been rigorously tested \cite{Smith2010}. As scientists, we should treat peer review like a drug and compare it with alternatives that might be more efficacious, cheaper, or with fewer side effects. Like the human body, the world of academia is complex and may react unexpectedly. For example, it surprised many people that application numbers decreased when the U.S. National Science Foundation removed funding deadlines \cite{NSF2016}.
 %Applicants in the new system were likely to submit their applications when ready, not when they were forced to do so by the deadline.

Besides peer review, bibliometrics is the most commonly used method for science evaluation. Although the method has broad acceptance in science \cite{reymert2021bibliometrics}, many criticisms have been published  \cite{macroberts2010problems}. Using evaluative bibliometrics may be improved if the indicators are theory-based, valid, adequately operationalized, standardized, and properly field-adjusted.
Just as medical operations should only be done by experienced physicians, evaluative bibliometrics should only be done by experienced, professional bibliometricians.
Advanced indicators include indicators that capture gaming of the system and additional indicators of good research practices (e.g., data and code sharing, protocol registrations) \cite{ioannidis2023defense, ioannidis2023quantitative}. Bibliometric data could be extended to develop merit indicators, such as indicators for independence \cite{van2019measuring}, or a score for the research topics covered by a researcher\cite{mom2023determinants}. Several new indicators and indicator variants have been developed in the field of bibliometrics that have been established for many years. 
Despite various efforts to improve evaluative bibliographics, there are several movements today against the (extensive) use of metrics (in national evaluation systems) such as the EU Agreement to Reform Research Assessment \cite{CoARA2023}. An overview of these movements can be found in \cite{rushforth2023practicing}. Movements are characterized by a preference for the peer review process and responsible use of metrics. Since both methods of evaluating research have their own strengths and weaknesses, a clear decision about one solution cannot be made in all evaluation situations. It is probably the best solution to design a specific evaluation framework for each task that combines both methods in most cases \cite{rushforth2023practicing, bornmann2019heuristics}. Since most completed evaluations do not investigate the success or failure of funding decisions - except the relatively small number of studies that focus on the predictive validity of grant evaluation (see section~2.1) - the effect of evaluations on science success remains unclear since the advent of these evaluations. Since many of the economic costs in evaluations are time investments (from various stakeholders, such as applicants and reviewers), investigations of peer review processes should especially focus on this issue. As bibliometric support may be a decisive element in reducing time investments, research is needed to understand how bibliometrics can play an optimal role in in grant decision-making \cite{van2020bibliometrically}. This should go hand in hand with selecting appropriate indicators and ways to improve their quality.

Using citations from peers worldwide as an ex-post evaluation procedure could offer an alternative to traditional ex-ante evaluation methods. While it would require time to accrue citations, endorsements from the relevant research community would add credibility to the decisions. Rather than relying on a small and often opaque group with unclear selection criteria, researchers would be assessed by the broader scientific community through their citation counts. This approach would democratize the evaluation process, ensuring that the direction of science is not dictated solely by a selected few. Ultimately, entrusting evaluation to the scientific community could be the most efficient and democratic solution, fostering transparency and inclusivity in the assessment of research projects \cite{ioannidis2023defense}.
\bigbreak

\textbf{\emph{Fourth: Alternative distribution systems.}} 

%Competition in science is not limited to the allocation of grants. Competition (as a process) should compete with other forms of distribution processes in grant funding. 
%In some alternative evaluation and distribution systems such as commissioned research, lottery, or base funding, part of the competition will disappear altogether and not just be shifted to another level. 

In many current systems, a large part of applications are investigator-led, meaning the investigator proposes what to study. 
Other types of funding are thematic grants, where themes are defined by the funding organizations, and commissioned research, where research question(s) are formulated by organizations or companies and investigators write applications explaining why they are qualified to answer the question(s). The potential benefit of thematic grants and commissioned research provides a more direct return to society than investigator-led, especially if the commissioning process is extensive and done in partnership with end users \cite{Chalmers2014}. These alternatives should never consume 100\% of a funder's budget, as investigator-led research is still a vitally important avenue for discovery. The ideal distribution of the budget over the different types of grants to spend on the different types of commissioned research is another unknown.

%In alternative, commissioned research, the research question(s) are created, and investigators write applications explaining why they are qualified to answer the question(s). The potential benefit is that commissioned research provides a more direct return to society than investigator-led, especially if the commissioning process is extensive and done in partnership with end users \cite{Chalmers2014}. This alternative should never consume 100\% of a funder's budget, as investigator-led research is still a vitally important avenue for discovery. Another unknown is the ideal share of the budget to spend on commissioned research.
% Adrian: the long list of problems with commissioned research could equally be applied to investigator-led
Commissioned research could fail if the process of selecting the topics is flawed and biased and leads to poor, conflicted, and/or dead-end investments. A problem with most current funding systems may be that even when research programs are not entirely commissioned, there is a strong hidden element of the commission behind them. For example, requests for applications may create boundaries around what ideas and topics are desirable, even if they seem to be relatively open and non-specific. 

Modified lotteries are an alternative system and one that has recently been trialed by funders in several countries, e.g., New Zealand, Switzerland, Germany, and the UK \cite{luebber2023rethink, stafford2023next,de2022modified, liu2020acceptability}. In a modified lottery, the applications are assessed by peer review. However, they are not ranked linearly, with a funding line splitting applications into funded and not funded. 
Instead, peer review is used to assess applications as fundable or not, with the winners drawn from those assessed as fundable. In a variation, reviewers can also select ``excellent'' proposals that should be funded without the risk of lottery. 
When peer review has been performed on some or all applications, another decision is whether to have weighting in the lottery (e.g., to give more tickets to applications with more favorable peer assessments) \cite{shaw2023a}.
Early results from The British Academy (the UK’s national academy for the humanities and the social sciences) suggests that the diversity of applicants increased after the modified lottery system was introduced, possibly because applicants outside the mainstream perceive the lottery system as fairer and so are more willing to apply \cite{academy}. 
The total cost of putting these applications together (and of reviewing them) is likely to depend on how strong a role peer review plays in the modified lottery, and how extensive proposals are reviewed. If peer review only consists in a minimal formal check for eligibility, the costs for reviewing and creating applications will be much lower than if traditional peer review is used.

An alternative distribution system that significantly reduces bureaucracy is to give every scientist a small amount of base funding \cite{Vaesen2017}. However, the problem then becomes how to define eligible scientists. 
For example, universities could switch the job description from technical staff to scientists to get more funding. This could be policed, but that would likely involve every scientist describing their work, which would then need to be peer reviewed, and hence the bureaucracy savings could be lost. 

A related system of base funding is to give money to each scientist but then force them to give away half \cite{bollen2017}. This uses the wisdom of the crowds to allocate more funding to those scientists who are respected by their peers. However, it could be prone to bullying as senior scientists may strong-arm junior scientists into promising them funding, with consequences if the promises are unfulfilled. However, we can only speculate on potential problems and benefits without properly testing the system.

A pilot investigation in Australia asked a large number of scientists to name up to ten scientists in the country who they thought would be deserving of funding \cite{barnett2017using}. The study showed that this democratic voting process reduced time requirements compared with a traditional grant review system. However, there was some bias in favor of naming scientists from the same institution, and some caveats were noted by participants regarding the potential for vote rigging, lobbying, and turning science into a popularity contest.  

In conclusion, we do not know what the best distribution system is.
A systematic review suggests that greater funding dispersal is likely to be beneficial \cite{aagaard2020concentration}. The best way to implement such greater dispersal and achieve maximum benefits remains unclear and needs careful evaluation. 
This uncertainty provides a solid basis for experimental tests.
Should fellowships come with additional funding, e.g., PhD stipends? What mix of the funding budget should be spent on early-career versus senior researchers? What mix of the funding budget should be spent on a project versus people funding? However, some of these questions could be answered using randomized trials, requiring researchers to consent to participating in experiments, which some will be unwilling to do. 

\bigbreak

\textbf{\emph{Fifth: Economic costs.}} The cost of competition (including positive external effects) as a function of the acceptance rate should be analyzed for different disciplines and along the continuum of applied and basic research. Therefore, we need data on (i) the time needed to prepare applications, (ii) possible positive indirect effects of proposal writing, (iii) the costs of decision-making processes including peer review and panel decisions, (iv) administrative costs of funding agencies and other organizations such as those involved in tendering for funding, (v) resources spent on training researchers to improve their grant proposal writing skills, including workshops, courses, and mentoring programs, and (vi) acceptance rates of the respective funding programs. 
When studying these economic costs, some specific issues should be considered. For example, it is essential to distinguish between the different evaluation and decision-making forms, as time investments may differ radically. 
To address the issue with self-reporting of time, studies are needed that analyze the discrepancies between the self-reported time estimates and a logged version of these data \cite{parry2021systematic}.

Another issue is that one should distinguish between the skill levels of applicants. Frequent applicants may become more efficient (and probably more successful) over time, indicating a learning effect. If this is the case and research should be done to find out, a division of labor may reduce the costs for grant writing: Not everyone has to be good at writing grants. Research is also needed to record who is actually doing the grant writing, e.g., it is possible that in many systems, much of the effort is contributed by managerial staff or professional grant writers rather than scientists. If so, this might mean that a large part of the devoted effort does not affect research resources, but it nevertheless entails costs.

\bigbreak

\textbf{\emph{Sixth: Epistemological costs.}}

In section 3.2, we distinguished between different aims of funding systems, including funding risky research, funding normal science, and funding proliferation and variety. Funding one of these goals is likely at the expense of the others and, therefore, a balance is needed. The optimization depends partly on understanding whether a relation exists between the evaluation and selection process and the level at which the specific aim is realized. Research on this has been scarce up to now. 
The relation should be studied at the level of the review and decision-making process of individual applications, but also at the level of the funding organization and its instruments. 

Only with hindsight can one assess whether disruptive projects have been funded \cite{tatsioni2010sources}, whether there has been enough variety in the funded research, and whether dead ends have been abandoned in time. Some research has suggested the conservative effects of competition and peer review as selection mechanisms, but that does not preclude that the funding instrument contributed meaningfully to one or more of the mentioned goals - as that cannot be evaluated at the individual project level. So the issue is to evaluate the outcomes of the funding instrument and funding organizations on the system level: Has one funding organization proved to be better at funding breakthroughs than another organization, and why? Do differences relate to how a funding organization defines evaluation criteria and organizes selection processes? The same holds for other epistemic aims, such as stimulating variety and selecting promising developments. 
Discussing these questions regarding the costs of competition also leads to the macro- (country) level. Are countries with a higher share of competitive funding less good at simultaneously supporting risky research, normal science, variety, and stimulating the promising parts of this variety? One may speculate that (too much) competition clearly entails epistemic costs.

%\begin{itemize}
%    \item Conservative effect of competition and peer review.
%    \item There are hypotheses that this effect exists, and there are simulations, but it would be good to somehow develop a method and measure it.
%   \item studies about overfunding and underfunding: some studies are calculating a 'funding sweet spot', and many studies show that competitive funding doesn't lead to an efficient distribution of funds. However, this research should be done for more fields and countries (not just biomed in the states).
%  \item tudies on depluralizing effects of the system in particular fields (e.g. less long term studies because grants are usually for a short period of time, less theoretical work and more applied research, ...)
%\end{itemize}

%\begin{itemize}
%    \item add this comment from Peter here: only with hindsight one can talk about 'disruptive', so not peer review but evaluation the outcomes of the funding institution is the way to go: does with hindsight a funder has proved to be better in funding breakthroughs than than another funder. 
%\end{itemize}

\bigbreak

\textbf{\emph{Seventh: Social and ethical costs.}} Our current understanding of the social and ethical costs associated with competitive grant funding has largely been derived from descriptive surveys and interviews. 
These methods remain valuable and could be used to obtain a more detailed insight into problems identified by existing research, such as the extent to which researchers return funds to funders or rapidly spend it before the grant expires, the extent to which funders check for negative and positive conflicts of interest, and the extent to which reviewers account for questionable and responsible research practices (e.g., preregistration) in evaluating grants. 
However, there is little research on the social costs incurred by individuals who miss out on grant competitions. For example, further exploration into whether the lack of funding for an early career initiates a negative spiral, potentially reducing mental health, motivation, and long-term productivity, could be interesting given how expensive and scarce research positions are.

However, most importantly, we should move beyond descriptive research and delve into the root causes of the social and ethical costs of science funding. Designing effective interventions without a clear understanding of the underlying causal structure is challenging. Researchers should aim to formulate causal hypotheses, assess their amenability to interventions, and test these hypotheses empirically through experiments in collaboration with funding organizations. 
Thus, as with many of the aspects of research funding which we discuss here, we believe that conducting more causal studies is paramount in uncovering actionable insights that can inform the policy and mechanism of science funding.

Another notable gap in our current knowledge lies in understanding the social and ethical costs associated with changes to the current system or alternative funding systems. 
It is particularly difficult to capture the complexity of these systems with simulations, and predicting outcomes is challenging. While single experiments have limitations (social and ethical costs may only arise when a funding method is broadly implemented), they remain an indispensable tool. Therefore, we call for increased experimentation with alternative funding models. Such studies should extend beyond assessing epistemic and financial implications; 
researchers should also investigate the impact on the well-being and questionable research practices of those involved, providing a comprehensive understanding of the broader consequences of alternative funding structures. For example, it may well be that funding groups rather than individuals positively impact the responsible research practices of researchers \cite{tiokhin2021shifting}. Using experiments, such hypotheses could be tested simultaneously with the epistemic benefits of such a system.
Table~\ref{tab} summarises our recommendations for a better understanding of the impact of competition on allocating scarce research funds.

\begin{table}[htbp]
    \centering
    \caption{Road to progress}
    \label{tab} 
    \begin{adjustbox}{max width=\linewidth}
        \includegraphics{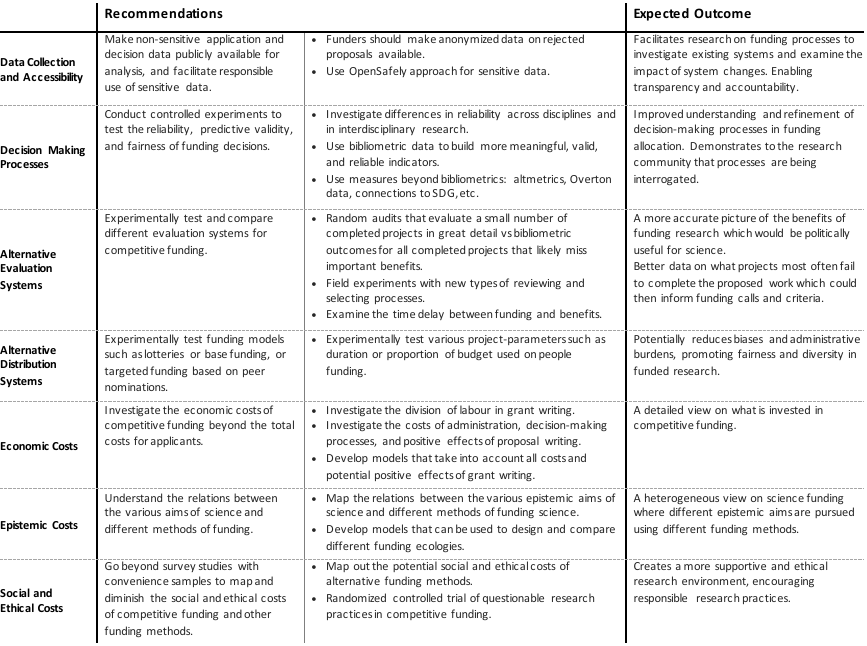}
    \end{adjustbox}
\end{table}

\bigbreak

\section*{Acknowledgments}
We thank Stephan Pühringer from Johannes Kepler University Linz and Ulf Heyman from Uppsala University for fruitful discussions.
The work of Stijn Conix for this paper was funded by the Fonds de la Recherche Scientifique—FNRS under grant no. T.0177.21. 
\printbibliography

\end{document}